\documentclass[prd,twocolumn,showpacs,showkeys,preprintnumbers,amsmath,amssymb,floatfix]{revtex4}

\usepackage[utf8]{inputenc}
\usepackage{hyperref}

\usepackage{dcolumn}% Align table columns on decimal point
\usepackage{bm}% bold math
\usepackage{graphicx}
\usepackage{subfigure}

\def\be{\begin{equation}}
\def\ee{\end{equation}}
\def\bestar{\begin{equation*}}
\def\eestar{\end{equation*}}
\begin{document}

\title{Shadow of a noncommutative inspired Einstein-Euler-Heisenberg black hole}

\author{Marco Maceda}
\email{mmac@xanum.uam.mx}
\affiliation{Departamento de F\'{i}sica \\ Universidad Aut\'{o}noma Metropolitana - Iztapalapa\\
                   A.P. 55-534, Ciudad de M\'{e}xico C.P. 09340, M\'{e}xico}         
                   
\author{Alfredo Macías}
\email{amac@xanum.uam.mx}
\affiliation{Departamento de F\'{i}sica \\ Universidad Aut\'{o}noma Metropolitana - Iztapalapa\\
                   A.P. 55-534, Ciudad de M\'{e}xico C.P. 09340, M\'exico}           
                   
\author{Daniel Martínez-Carbajal}
\email{danielmc@xanum.uam.mx}
\affiliation{Tecnol\'{o}gico de Estudios Superiores\\ del Oriente del Estado de M\'{e}xico\\
                   Paraje San Isidro s/n, col. Barrio de Tecamachalco, A.P. 41, Estado de  M\'{e}xico C.P. 56400,  M\'{e}xico}                 
                                 
\date{\today}

\begin{abstract}
Using a noncommutative inspired Einstein-Euler-Heisenberg black hole, we analyse the associated graviton and light rings that together with the event horizon, provide the necessary components to build its shadow. An analysis of the angular radius of the noncommutative inspired Einstein-Euler-Heisenberg black hole shows that for observers located in its vicinity, it becomes smaller when compared with the commutative case; this effect is more noticeable as noncommutativity increases. The existence of marginally stable bound orbits and the critical angle for the escaping of photons of a noncommutative inspired Einstein-Euler-Heisenberg star is also discussed.
\end{abstract}



\maketitle

%*****************
\section{Introduction}

The Reissner-Nordström (RN) solution~\cite{Reissner:1916,Nordstrom:1918} describes a static charged black hole, where the electromagnetic field comes from standard Maxwell electrodynamics. Nevertheless, we expect that for strong electromagnetic fields, this description will fail due to non-linear effects arising from the electromagnetic field. Besides, since a black hole has a singularity as a distinctive characteristic, quantum effects may happen quite naturally. As a consequence, to consider black hole solutions in General Relativity (GR) with non-linear electrodynamics super seeding Maxwell electrodynamics seems a logical path to follow.

Several non-linear electrodynamics are known~\cite{Born:1934gh,Bardeen:1968,AyonBeato:1998ub,Hendi:2014xia,Balart:2014cga}. Perhaps, one of the simplest is Euler-Heisenberg non-linear electrodynamics~\cite{Heisenberg:1935qt}; it arises quite naturally from calculations in Quantum Electrodynamics (QED) at one loop as the effective description of the vacuum as a fluctuating medium~\cite{Gravejat:2018}. It is natural to study Euler-Heisenberg (EH) non-linear electrodynamics when extending Maxwell electrodynamics due to its sound physical basis and experimental verification; it is relevant, for example, to describe magnetars~\cite{Baring:2000cr,Denisov:2003ba}.

A different source of quantum effects appears if we want to eliminate the singularity in solutions in standard GR. Since the standard notion of spacetime breaks down at the Planck scale, noncommutative geometry becomes the natural framework to analyse the dynamics at such short distances. For this purpose, smeared noncommutative distributions of mass and charge~\cite{Nicolini:2005gy,Nicolini:2008aj,Banerjee:2009xx} inspired from results in noncommutative quantum field theory~\cite{Smailagic:2003rp,Smailagic:2004yy} are used. This approach was employed to construct the noncommutative inspired Schwarzschild, RN, Kerr and Kerr-Newman~\cite{Nicolini:2005vd,Modesto:2010rv}; extensions to incorporate non-linear electrodynamics exist~\cite{Gonzalez:2014mza,Gonzalez:2015mpa,Maceda:2018zim}. The principal outcome is that the singularity at the location of the source disappears, and instead a regular black hole solution may exist.

On the other hand, the shadow of a black hole serves as a probe of the very nature of the black hole~\cite{Synge:1966,Bardeen:1973tla,Falcke:1996dk,Falcke:1999pj,Grenzebach:2014fha,Perlick:2015vta,Grenzebach:2017fnp,Perlick:2018iye,Dymnikova:2019vuz,Creci:2020mfg}; it is also useful to understand how it interacts with its surroundings. As it is well-known, the shadow depends on the physical parameters of the black hole such as its mass, charge and angular momentum. Even though a complete description of all the processes occurring in the vicinity of the black hole requires at the end of a numerical approach, it is also true that a theoretical analysis sheds insight into the significant or relevant properties that are present in astronomical observations. The recent direct observation of the shadow of a black hole in the M87* galaxy by the Event Horizon Telescope collaboration~\cite{Akiyama:2019cqa} makes this analysis more compelling to pursue.

In the case of noncommutative inspired black holes, the shadow becomes the right tool to investigate the quantum nature of spacetime. In particular, the shadow for rotating noncommutative inspired black holes with Maxwell electrodynamics shows a significant distortion from the classical case due to modifications on the dragging velocity for co-rotating and counter-rotating photon orbits~\cite{Sharif:2016znp}. For some non-linear electrodynamics results exist for the static~\cite{Saha:2018zas,Maceda:2018zim}; even in standard GR, generating rotating black hole solutions with non-linear electrodynamics is still a difficult task~\cite{CiriloLombardo:2004qw,CiriloLombardo:2006ph,Breton:2019arv}.

In this paper, we want to gain further insight into the properties of the regular noncommutative inspired black holes in the presence of non-linear electrodynamics. For this purpose, we will analyse the paths followed by gravitons in the background of the static noncommutative inspired Einstein-Euler-Heisenberg (nciEEH) black hole to determine the corresponding shadow of the black hole in detail. Since photons travelling in a curved spacetime coupled to non-linear electrodynamics follow trajectories that are determined by the Plebanski metric $\gamma_{\mu\nu}$, we will use this metric to analyse the shadow. In this way, we shall infer if features of the structure of the spacetime at a quantum scale may be susceptible to observation.

We organise this paper as follows: In Sec.~\ref{secc:2}, we briefly review some features of the static nciEEH black hole solution and define its noncommutative screening charge. In Sec.~\ref{secc:3}, we analyse circular and marginally stable orbits of test particles restricted to the equatorial plane; we obtain then the critical semi-angle for the emission (absorption) of photons from the surface of a nciEEH star in Sec.~\ref{secc:4}. Afterwards, we focus on graviton and light rings in Sec.~\ref{secc:5} to determine the required elements to build the shadow of the nciEEH black hole; the condition for the existence of a light ring is expressed in terms of its noncommutative screening charge. We also analyse the angular radius of the shadow of the nciEEH in this section. Finally, we end with some comments and perspectives in the Conclusions section. We use geometric units $G=1=c$ throughout the paper.

%*****************
\section{Screening charge of the nciEEH black hole}
\label{secc:2}

In~\cite{Maceda:2018zim}, the static nciEEH black hole was defined using a spherical symmetric metric
\be
ds^2 = - f(r) dt^2 + f(r)^{-1} dr^2 + r^2 d\vartheta^2 + r^2 \sin^2 \vartheta d\phi^2,
\label{ds2}
\ee
where $r, \vartheta$ and $\phi$ have the usual meanings; for the electrical charged solution, the metric function $f(r)$ is~\cite{Maceda:2018zim} 
\begin{eqnarray}
f &=& 1 - \frac {4M}{r\sqrt{\pi}} \gamma\left( \frac 32, \frac {r^2}{4\theta} \right) + \frac 1\pi \frac {Q_e^2}{r^2} \gamma^2 \left( \frac 12, \frac {r^2}{4\theta} \right) 
\nonumber \\[4pt]
&& + \frac 1\pi \frac {Q_e^2}{r^2} \left[ \sqrt{\frac 2\theta} r \gamma \left( \frac 32, \frac {r^2}{4\theta} \right) - \frac r{\sqrt{2\theta}} \gamma \left( \frac 12, \frac {r^2}{2\theta} \right)  \right]  
\nonumber \\[4pt]
&&- \frac {4A}{\pi^2} \frac {Q_e^4}r \int_r^\infty \frac {ds}{s^6} \gamma^4 \left( \frac 32, \frac {s^2}{4\theta} \right)
\nonumber \\[4pt]
&&+ \frac {A}{8\pi^2} \frac {Q_e^4}r 
\left[ 1- \frac 2{\sqrt{\pi}} \gamma \left( \frac 32, \frac {r^2}{4\theta} \right) \right] \frac {c_0}{\theta^{5/2}},
\label{dsnceeh}
\end{eqnarray}
where $M$ is the ADM mass~\cite{Nicolini:2008aj,Modesto:2010rv}, $\theta$ is the noncommutative parameter and $\gamma(n,z)$ is the lower incomplete gamma function~\cite{Abramowitz:1965}; the presence of this function in the metric points to the use of a smeared distribution of mass in the field equations instead of a point-like source and that the solution is non-perturbative on $\theta$. $A$ is a constant of the model that measures the non-linearity of the EH electrodynamics, but in the context of QED, it has a very well-defined value~\cite{Heisenberg:1935qt}; finally, $c_0 := \int_0^\infty ds \,s^{-6} \gamma^4 \left( \frac 32, s^2 \right) = 0.02757$. 

\begin{figure}[htbp]
\begin{center}
\includegraphics[width=8cm]{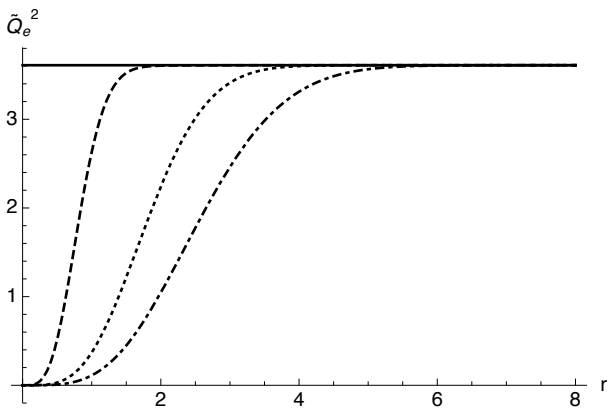}
\caption{Screening noncommutative charge $(\tilde Q_e)^2$ for $\theta = 0.1, 0.5$ and $1$ (dashed, dotted and dash-dotted lines respectively) with $Q_e = 1.9, A = 0.1$; the solid horizontal line corresponds to the constant commutative value $Q_e^2 = 3.61$.}
\label{fig1}
\end{center}
\end{figure}

The electromagnetic content of the electrical charged solution is given by the Plebanski variables~\cite{Plebanski:1970}
\be
P_{\mu\nu} = \frac 2{\sqrt{\pi}} \frac {Q_e}{r^2} \gamma \left( \frac 32, \frac {r^2}{4\theta} \right) \delta^t_{[\mu} \delta^r_{\nu]}. 
\ee 
It follows that the electromagnetic Lorentz invariants in terms of the Plebanski variables are then~\cite{Maceda:2018zim} $s = \frac 12 P_{01}^2, t = 0$. 
Let us define the following noncommutative quantities 
\begin{eqnarray}
\tilde M &:=& \frac {2 M}{\sqrt{\pi}} \gamma \left( \frac 32, \frac {r^2}{4\theta} \right) \quad \stackrel{\theta \to 0}{\longrightarrow} \quad M,
\nonumber \\[4pt]
(\hat Q_e^{nc})^2 &:=& \frac {Q_e^2}\pi \left[ \gamma^2 \left( \frac 12, \frac {r^2}{4\theta} \right) + \sqrt{\frac 2\theta} r \gamma \left( \frac 32, \frac {r^2}{4\theta} \right) \right.
\nonumber \\[4pt]
&&\left.- \frac r{\sqrt{2\theta}} \gamma \left( \frac 12, \frac {r^2}{2\theta} \right) \right] \stackrel{\theta \to 0}{\longrightarrow} \quad Q_e^2.
\end{eqnarray}
Using the above definitions, we write then the function $f$ in its RN-like form as
\be
f = 1 -\frac {2\tilde M}r + \frac {(\tilde Q_e)^2}{r^2},
\label{ds2rn}
\ee
where
\begin{eqnarray}
&&(\tilde Q_e)^2 := (\hat Q_e^{nc})^2 
\nonumber \\[4pt]
&&\times \left\{ 1 - \frac {4Ar}{\pi^2} \frac {Q_e^4}{(\hat Q_e^{nc})^2} \left[ \displaystyle\int_r^\infty \frac {ds}{s^6} \gamma^4 \left( \frac 32, \frac {s^2}{4\theta} \right) \right.\right.
\nonumber \\[4pt]
&&\left.\left.- \frac 1{32} \left[ 1- \frac 2{\sqrt{\pi}} \gamma \left( \frac 32, \frac {r^2}{4\theta} \right) \right] \frac \alpha{\theta^{5/2}} \right] \right\}.
\end{eqnarray}
The charge $\tilde Q_e$ is seen as a noncommutative screening charge. Indeed, when $\theta \to 0$ or equivalently $4\theta \ll r^2$, $(\tilde Q_e)^2$ reduces to the screening charge defined in~\cite{Ruffini:2013hia} for the commutative case. In a similar way,  $\tilde M$ may be interpreted as a sort of noncommutative screening mass since $\tilde M \approx M(1 - \frac r{\sqrt{\theta\pi}} e^{-r^2/4\theta})$ for $4\theta \ll r^2$, and it is only for an observer located at spatial infinity that it becomes the ADM mass; nevertheless, this analogy does not go further than that. We also immediately obtain from Eq.~(\ref{ds2rn}) that the horizon radius is such that
\be
r_h = \tilde M \pm \sqrt {\tilde M^2 - (\tilde Q_e)^2},
\ee
and that the extremal configuration is determined by $\tilde M = |\tilde Q_e|$; even though not so immediate, we may show that due to the screening, extremal configurations exist where $M < Q_e$ for the nciEEH black hole. For illustrative purposes, we show in Fig.~\ref{fig1} the typical behaviour of $\tilde Q_e$ as a function of $r$: for small values of $r$ the noncommutative screening charge differs appreciably from the constant commutative value and it is only for large values of $r$, or equivalently when $4\theta \ll r^2$ (small noncommutativity), that it becomes constant. 

%******************
\section{Circular and marginally stable bound orbits}
\label{secc:3}

\begin{figure}[htbp]
\begin{center}
\includegraphics[width=8cm]{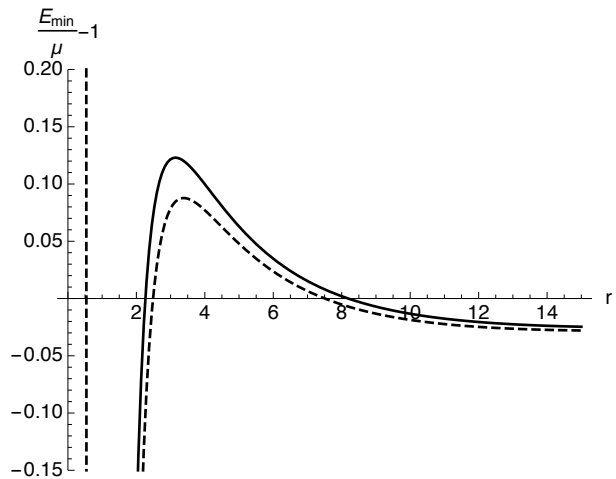}
\caption{Typical behaviour of $E_{min}$ as a function of $r$ for $\theta = 0$ (solid line) and $\theta = 0.1$ (dashed line) with $A=0.8$; notice that there exist unstable, stable and marginally stable circular orbits.}
\label{fig2}
\end{center}
\end{figure}

We start by discussing general facts about the motion of test particles in the gravitational field of a spherical symmetric noncommutative inspired black hole with line element as given in Eq.~(\ref{ds2}). The function $f$ defines the noncommutative inspired black hole; it depends on the radial variable $r$ and the noncommutative parameter $\theta$. The Lagrangian approach~\cite{Papapetrou:1974gq,Landau:1979tf,Chandrasekhar:2009} to analyse the orbital motion of test particles in curved backgrounds tells us then that from
\be
2{\cal L} = - f \dot t^2 + f^{-1} \dot r^2 + r^2 \dot \vartheta^2 + r^2 \sin^2 \vartheta \dot \phi^2,
\label{lagrangian}
\ee 
we have two conserved quantities
\be
E := p_t = - f \dot t, \qquad L:= p_\phi = r^2 \sin^2 \vartheta\, \dot \phi,
\ee
corresponding to energy and angular momentum respectively, together with a first integral of motion given by
\be
\dot r^2 + f r^2 \dot \vartheta^2 + \frac {L^2 f - E^2 r^2 \sin^2 \vartheta}{r^2 \sin^2 \vartheta} = 2 f {\cal L}.
\ee
For massive test particles with rest mass $\mu$, we set ${\cal L} = -\frac 12 \mu^2$ and $\vartheta = \pi/2, \dot \vartheta = 0$. The last two conditions restrict us to equatorial orbits; due to the spherical symmetry we can consider only these orbits without loss of generality~\cite{Chandrasekhar:2009}. We have then
\be
\left( \frac {dr}{d\tau} \right)^2 = E^2 - \frac f{r^2} L^2 - \mu^2 f =: \frac 1{r^2} R. 
\ee
Once a radius $r$ and angular momentum $L$ is specified, the values of $E$ that are physically meaningful lie above a minimum value $E_{min}$; the condition $R=0$ fixes this value to be
\be
\frac {E_{min}}\mu = \frac {(r^2 f)^{1/2}}{r^3} \left( r^2 \frac {L^2}{\mu^2} + r^4 \right)^{1/2}. 
\ee
For the case of the noncommutative inspired EEH black hole with $f$ given by Eq.~(\ref{dsnceeh}), we notice that since $f$ is asymptotically flat we have $E_{min}/\mu \to 1$ for large values of $r$. Also, stable, bound circular orbits exist as well as unstable unbound circular orbits as shown in Fig.~\ref{fig2}; they correspond to a minimum or a maximum of $E_{min}$ respectively. We see that an unstable orbit in the presence of noncommutativity requires a smaller value of $E_{min}$ than in the commutative case.

In the case of a massive test particle with energy $E$ and angular momentum $L$, the conditions for circular orbits are $R =0, R_{,r} = 0$; we have then
\begin{eqnarray}
r^2 E^2 - f L^2 - \mu^2 r^2 f = 0,
\nonumber \\[4pt]
2r E^2 - f_{,r} L^2 - 2\mu^2 r f - \mu^2 r^2 f_{,r} = 0.
\end{eqnarray}
It follows that
\begin{eqnarray}
\frac E\mu = \frac {2^{1/2} f}{(2f - r f_{,r})^{1/2}},
\qquad 
\frac L\mu = \frac {(r^3 f_{,r})^{1/2}}{(2f - r f_{,r})^{1/2}}.
\end{eqnarray}
From the first relation, we see that the binding energy~\cite{Bardeen:1973tla}
\be
E_{binding} := 1 - \frac E\mu,
\ee
possesses a critical point when
\be
3 f f_{,r} - 2 r f_{,r}^2 + rf f_{,rr} |_{r_{ms}} = 0.
\ee
The value $r = r_{ms}$, that is a solution to the above equation, determines the radius of a marginally stable bound orbit. 

%**********
\section{Critical semi-angle of emission for photons in a nciEEH star}
\label{secc:4}

\begin{figure}[htbp]
\begin{center}
\includegraphics[width=8cm]{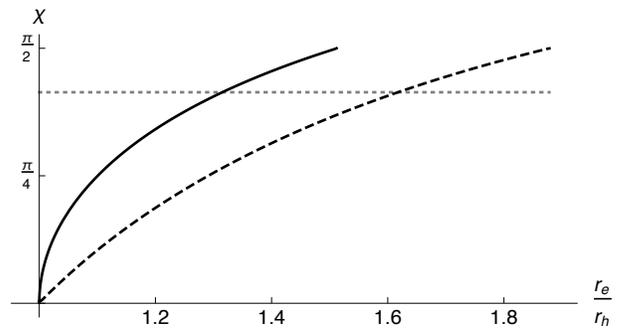}
\caption{Critical semi-angle $\chi$ for emission of photons in the commutative (solid line) and noncommutative ($\theta = 0.237$, dashed line) spacetimes for a nciEEH star ($A=1$).}
\label{fig3}
\end{center}
\end{figure}

The following discussion may be applied to any star of mass $M$ emitting photons and whose exterior gravitational field is described by the metric in Eq.~(\ref{ds2}). As in the previous section, we first notice that there are two conserved quantities that we write now as
\be
r^2 \dot \vartheta = \eta, \qquad f \dot t = \eta \sigma,
\ee
where the dot means derivation with respect to an affine parameter. For null geodesics with $\phi = const.$, we obtain then
\be
\dot r^2 = \eta^2 \left( \sigma^2 - \frac f{r^2} \right),
\ee
where $r \in [r_h, \infty)$ with $f(r_h) = 0$; it follows that
\be
\left( \frac {dr}{d\vartheta} \right)^2 = r^4 \left( \sigma^2 - \frac f{r^2} \right).
\ee
When a photon is emitted from the surface of a star, the inclination angle $\psi \in [0, \pi/2]$ of the ray with respect to the radial direction is then~\cite{Synge:1966}
\be
\cot ^2 \psi = f^{-1} r^{-2} \left( \frac {dr}{d\vartheta} \right)^2 = r^2 f^{-1} \sigma^2 - 1 =: r^4 F(r).
\ee
On the other hand, given initial conditions $(r_e, \psi_e)$ for emission, we obtain
\be
\sin^2 \psi_e = \frac {f(r_e)}{r_e^2 \sigma^2} \qquad \Rightarrow \qquad \sigma^2 = \frac {f(r_e)}{r_e^2} \csc^2 \psi_e. 
\ee
The condition $F = 0$ indicates the presence of an apse for the ray and in consequence, it falls back into the star; it also implies that $\sigma^2 = f/r^2$. As a function of $r$, $\sigma^2$ attains a maximum at $r = r_{max}$ given by 
\be
r_{max} f_{,r} (r_{max}) = 2 f(r_{max}). 
\ee
The condition $\sigma^2 \geq \sigma_{max}^2 $ guarantees that photons escape to infinity; we have thus
\be
\sin^2 \psi_e \leq \frac {r_{max}^2}{f(r_{max})} \frac {f(r_e)}{r_e^2}.
\ee
The critical semi-angle $\chi$ for escaping is then
\be
\sin^2 \chi = \frac {r_{max}^2}{f(r_{max})} \frac {f(r_e)}{r_e^2}.
\ee
In Fig.~\ref{fig3} we illustrate the behaviour of $\chi$ for both the commutative and noncommutative cases. We first notice that in the gravitational strong limit, $r_e = r_h$, the critical semi-angle of emission vanishes and thus, as it happens in the commutative case, only rays emitted in the direction normal to the surface of the star escape. Nevertheless, we also see that for a given value of $\chi$, the ratio $r_e/r_h$ is larger in the noncommutative case when compared to the standard case. We may interpret this result as saying that since we have a smeared distribution of mass instead of a point-like distribution for the nciEEH star, we need a larger sphere for emission to produce the same effect as in the commutative case; if observable, this effect may provide us with a direct evidence of noncommutativity.  

%**********
\section{Graviton, light rings and shadow of the nciEEH black hole}
\label{secc:5}

We now proceed to determine the shadow of the black hole using the Lagrangian approach. First, we consider the trajectories followed by massless tests particles; a candidate for this kind of particles is the graviton due to the constraints in its mass~\cite{LIGOScientific:2018mvr,LIGOScientific:2019fpa,Bernus:2019rgl} and in the following, we will refer to them in this way. Using Eq.~(\ref{lagrangian}) with $\vartheta = \pi/2, \dot \vartheta = 0$, then the condition ${\cal L} = 0$ restricts us to equatorial orbits followed by gravitons; due to spherical symmetry, we may analyse only these orbits without loss in generality. With these choices, we obtain an effective potential 
\be
V_{eff}(r) = L^2 \frac {f(r)}{r^2} - E^2,
\ee
to describe the graviton orbits in terms of their angular momentum $L$ and energy $E$. Important constituents of the shadow of the black hole are the circular graviton orbits at a certain radius $r_{gr}$; they are defined by the conditions $V_{eff} (r_{gr}) = 0 = V_{eff,r} (r_{gr})$. In our case, they become
\be
b_{gr}^2 := \frac {L^2}{E^2} = \frac {r_{gr}^2}{f(r_{gr})}, \qquad r_{gr} f_{,r} (r_{gr}) - 2 f(r_{gr}) = 0.
\label{brgr}
\ee
$b_{gr}$ is the impact parameter of the graviton orbit; gravitons inside a sphere of radius $b_{gr}$ get trapped by the black hole and, if unperturbed, can not escape to a distant observer. Notice that $b_{gr}^2$ is essentially the metric coefficient $g_{\phi\phi} = r^2 \sin^2\theta$ upon the metric coefficient $-g_{tt} = f(r) $ of the line element Eq.~(\ref{ds2}), both being evaluated at $r = r_{gr}, \theta = \pi/2$.

We now apply the above results to the nciEEH black hole using its RN-like form; we deduce then that the graviton ring are located at
\begin{eqnarray}
&&2 (1 + \tilde M_{,r}) r_{gr} = 3\tilde M + \tilde Q_e \tilde Q_{e,r} 
\nonumber \\[4pt]
&&\pm \sqrt{(3\tilde M + \tilde Q_e \tilde Q_{e,r})^2 - 8(1 + \tilde M_{,r}) (\tilde Q_e)^2},
\end{eqnarray}
where $\tilde M = \tilde M(r_{gr}), \tilde Q_e = \tilde Q_e(r_{gr})$. Notice that in the commutative limit and in the absence of non-linearity, when $\tilde M \to M$ and $\tilde Q_e \to Q_e$, we recover the well-known result $2 r_{gr} = 3 M \pm \sqrt{9 M^2 - 8 Q_e^2}$ for the RN black hole.

\begin{figure}[htbp]
\begin{center}
\includegraphics[width=8cm]{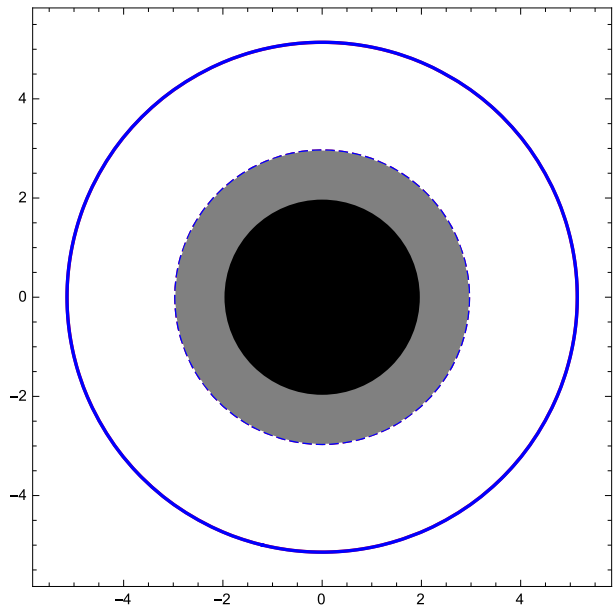}
\caption{Shadow of a nciEEH black hole with $M = 1, Q_e = 0.25, A = 0.5$ and $\theta = 0.1$. The (black) inner circle corresponds to the event horizon, the (gray) intermediate circle defines the graviton and light rings while the exterior circle corresponds to the impact parameters for the graviton and light rings.}
\label{fig4}
\end{center}
\end{figure}

The second main constituent of the shadow of a black hole is the light ring. We use the Plebanski metric~\cite{Plebanski:1970} $\gamma_{\mu\nu}$ since it indeed determines the trajectories followed by photons in a gravitational background coupled to a non-linear electrodynamics; in general we have that
\be
\gamma_{\mu\nu} := g_{\mu\nu} - 4\pi T^{e.m.}_{\mu\nu}. 
\ee
Applying the above definition to the electrically charged nciEEH black hole, we obtain
\begin{eqnarray}
&&\gamma_{tt} = g_{tt} (1 + A s), \qquad \gamma_{rr} = g_{rr} (1 + A s),
\nonumber \\[4pt]
&&\gamma_{\theta\theta} = g_{\theta\theta} (1 - A s), \qquad \gamma_{\phi\phi} = g_{\phi\phi} (1 - A s).
\label{gammag}
\end{eqnarray}
In consequence, by following the same argument as for the graviton orbit, we immediately deduce that the impact parameter for the light ring is
\be
b^2_{lr} = \frac {1 - A s(r_{lr})}{1 + A s(r_{lr})} \times \frac {r_{lr}^2}{f(r_{lr})} = [1 - 2 A s(r_{lr})] \times \frac {r_{lr}^2}{f(r_{lr})},
\label{impblr}
\ee
where $r_{lr}$ is a solution of 
\be
- \gamma_{tt} (\gamma_{rr} \gamma_{\phi\phi})_{,r} +  \gamma_{\phi\phi} (\gamma_{tt} \gamma_{rr})_{,r} \Big|_{r=r_{lr}, \theta = \pi/2} = 0.
\ee
Using the relations in Eq.~(\ref{gammag}), the previous condition becomes
\be
r_{lr} f_{,r} (r_{lr}) - 2 f(r_{lr}) + 2 A r_{lr} f(r_{lr}) s_{,r} (r_{lr}) = 0,
\label{condlr}
\ee
or equivalently
\begin{eqnarray}
&&2 (1 + \tilde M_{,r}) r_{lr} =  3\tilde M + \tilde Q_e \tilde Q_{e,r} 
\nonumber \\[4pt]
&&\pm \left[ (3\tilde M + \tilde Q_e \tilde Q_{e,r})^2 - 4 (1 + \tilde M_{,r}) \{ 2(\tilde Q_e)^2 \right.
\nonumber \\[4pt]
&&\left. - A r_{lr} [r_{lr}^2 - 2 \tilde M r_{lr} + (\tilde Q_e)^2 ] \} \right]^{1/2},
\end{eqnarray}
where in a similar way as before, $\tilde M = \tilde M(r_{lr}), \tilde Q_e = \tilde Q_e(r_{lr})$ and $\tilde Q_{e,r} = \tilde Q_{e,r} (r_{lr})$. Using all these results, we illustrate the graviton orbit, light ring and event horizon for a generic nciEEH black hole in Fig.~\ref{fig4}. 

Additionally, noncommutative corrections to the classical shadow can be determined in the limit of small noncommutativity ($4\theta \ll r^2$); to lowest order, we obtain the relation
\be
b_{lr}^2 = b_{lr}^{(0)\,2} + \delta b_{lr}^2 e^{-r_{lr}^{(0)\,2}/4\theta} +\dots,
\ee 
where $r_{lr}^{(0)}$ is the commutative light ring radius and $b_{lr}^{(0)}$ is the commutative impact parameter for the light ring of the commutative EEH black hole; we give the explicit expression for the coefficient $\delta b_{lr}^2$ in the Appendix. We notice that corrections are suppressed exponentially in this limit, reflecting the fact that the nciEEH black hole rapidly becomes the classical EEH black hole.

It is important to remark that light is the sole responsible for the shadow of a black hole; gravitons will not be visible. Indeed, a complementary calculation using the Hamilton-Jacobi approach shows that for light following the trajectories determined by the Plebanski metric $\gamma_{\mu\nu}$, the celestial coordinates~\cite{Bardeen:1973tla}
\begin{eqnarray}
\alpha &=& \lim_{r_0 \to \infty} \left( - r_0^2 \sin \vartheta_0 \frac {d\phi}{dr} \Big|_{\vartheta = \vartheta_0} \right),
\nonumber \\[4pt]
\beta &=& \lim_{r_0 \to \infty} \left( r_0^2 \frac {d\vartheta}{dr} \Big|_{\vartheta = \vartheta_0} \right),
\end{eqnarray}
where $(r_0, \vartheta_0)$ is the position of a distant observer, satisfy the relation
\be
\alpha^2 + \beta^2 = b^2_{lr}.
\ee
Physically, every photon emitted from the photon sphere becomes a definite point in the $(\alpha, \beta)$-plane for the distant observer and therefore, it becomes observable. 

\begin{figure}[htbp]
\begin{center}
\includegraphics[width=8cm]{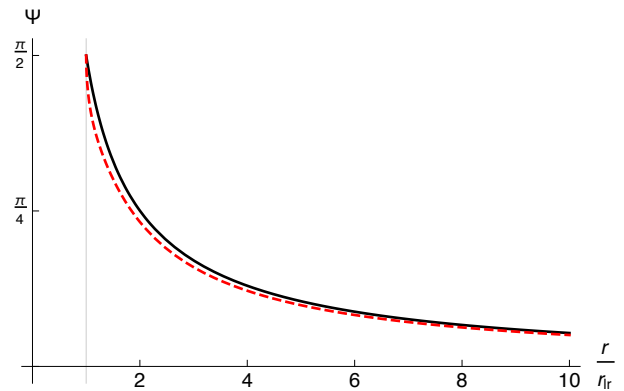}
\caption{Angular radius $\Psi$ as a function of the distance from the photon sphere to the observer for the nciEEH black hole. The solid line corresponds to $\theta = 0.01$ (weak noncommutativity) and the dashed line to $\theta = 0.6$ (strong noncommutativity); in both cases $A = 0.5$.}
\label{fig5}
\end{center}
\end{figure}

On the other hand, astrometric observables are also useful tools to analyse the shadow of a black hole~\cite{Chang:2020miq}; for static spherically symmetric space-times, the relevant quantity is the angular radius $\Psi$ of the shadow between a straight ray light from the center of the black hole and a light ray originating from the photon sphere. In Fig.~\ref{fig5}, we compare the behaviour of $\Psi$ for different values of the noncommutative parameter $\theta$. Notice that for large distances, distant observers measure the same value for $\Psi$ for weak and strong noncommutativity; this may be expected since for large distances, where $4\theta \ll r^2$, nciEEH black holes with different values of $\theta$ become essentially the same commutative one. On the other hand, observers located near the nciEEH black hole measure a smaller angular radius $\Psi$ as noncommutativity increases. 

%************************
\section{Conclusions}

For the nciEEH black hole, we have defined screening mass and charge quantities similar to those used in the commutative case; we notice that if $\theta \neq 0$, the screening charge is almost vanishing near the origin, increasing monotonically to the constant commutative value for great distances. 

We also showed that there exist stable and unstable orbits for the nciEEH black hole; moreover, marginally stable orbits are present just as in the commutative case. A complete analysis should also include the accretion disk surrounding the black hole along the lines of~\cite{Lahiri:2019mwc}. Another straightforward application of the nciEEH solution was the emission of photons from the surface of a star with an exterior gravitational field given by it. We showed that a less gravitational intense nciEEH star produces the same outgoing (incoming) semi-angle of emission (absorption) for photons emitted (absorbed) from its surface when compared with a commutative EEH star; this effect may be susceptible to observation in the future.

As mentioned in the Introduction, the shadow of a black hole contains relevant information about its nature. We constructed the shadow of a nciEEH black hole by taking into account the existence of light rings determined by an impact parameter $b_{lr}$; due to spherical symmetry, we indeed have a photon sphere surrounding the black hole. Inside the photon sphere, we find the graviton sphere and the event horizon, both not visible to outside observers. Differences exist between the nciEEH black hole and the RN black hole, but they are not quite striking at the level of the shadow. Nevertheless, they may be recognised when using the angular radius $\Psi$; a visible difference in the angular size of the nciEEH black hole will appear to observers situated in its close neighbourhood. 

%*********
\acknowledgments

We thank the President of the Universidad Autónoma Metropolitana for financial support through the Programa Especial de Apoyo a la Investigación (Project I13).

\appendix*

\section{Noncommutative lowest order corrections to the impact parameter of light rings}

In the limit of small noncommutativity, $4\theta \ll r^2$, the metric function $f(r)$ and the electromagnetic invariant $s(r)$ have the approximate expressions
\begin{eqnarray}
f &=& f^{(0)} + \delta f e^{-r^2/4\theta} + \dots,
\nonumber \\[4pt]
s &=& s^{(0)} + \delta s e^{-r^2/4\theta} + \dots,
\end{eqnarray}
where $f^{(0)}$ and $s^{(0)}$ denote the classical commutative quantities meanwhile $\delta f$ and $\delta s$ are the noncommutative corrections; for the nciEEH black hole, their explicit forms are as follow
\begin{eqnarray}
f^{(0)} &=& 1 - \frac {2M}r + \frac {Q_e^2}{r^2} - \frac A{20} \frac {Q_e^4}{r^6,}
\nonumber \\[4pt]
s^{(0)} &=& \frac {Q_e^2}{2r^4},
\nonumber \\[4pt]
\delta f &=& \frac {2M}{\sqrt{\pi \theta}} - \frac {4Q_e^2}{r^3} \sqrt{\frac \theta\pi} - \frac {Q_e^2}{\sqrt{2}\pi \theta} + \frac {A Q_e^4 c_0}{8 \pi^{5/2} \theta^3} 
\nonumber \\[4pt]
&&+ \frac {2AQ_e^4}{r^7} \sqrt{\frac \theta\pi},
\nonumber \\[4pt]
\delta s &=&-\frac {Q_e^2}{\sqrt{\pi \theta}} \frac 1{r^3}.
\end{eqnarray}
If we are interested in the leading noncommutative corrections to the impact parameter of light rings for the nciEEH black hole, we need the lowest order correction to the light ring radius $r_{lr}$. For this purpose, we write 
\be
r_{lr} = r_{lr}^{(0)} + \delta r_{lr} e^{-r_{lr}^{(0)2}/4\theta} + \dots,
\ee 
where $r_{lr}^{(0)}$ denotes its commutative value and substitute this expression in Eq.~(\ref{condlr}); we find that the noncommutative correction $\delta r_{lr}$ is
\be
\delta r_{lr} = \frac {q_1}{q_2} \Big|_{r = r_{lr}^{(0)}},
\ee
where
\begin{eqnarray}
q_1&:=&2\delta f - r \delta f_{,r} - 2 A r s^{(0)} \delta f - 2 A r f^{(0)} \delta s,
\nonumber \\[4pt]
q_2&:=&-f^{(0)}_{,r} + r f^{(0)}_{,rr} + 2 A f^{(0)} s^{(0)} + 2 A r f^{(0)}_{,r} s^{(0)} 
\nonumber \\[4pt]
&&+ 2 A r f^{(0)} s^{(0)}_{,r},
\end{eqnarray}
with $f^{(0)}_{,r}$ and $\delta f_{,r}$ defined by the relation
\be
f_{,r} = f^{(0)}_{,r} + \delta f_{,r} e^{-r_{lr}^{(0)2}/4\theta} + \dots.
\ee
Similar definitions apply for the calculation of $f^{(0)}_{,rr}$ and $s^{(0)}_{,r}$. With all these elements, we obtain from Eq.~(\ref{impblr}) the noncommutative correction to the impact parameter for the light rings in the form $\delta b_{lr}^2 e^{-r_{lr}^{(0)\,2}/4\theta}$ where
\begin{eqnarray}
\delta b_{lr}^2 &=& -b_{lr}^{(0)\,2} \left[ r_{lr}^{(0)} \left( f^{(0)}_{,r} \delta r_{lr} + \delta f \right) + 2 f^{(0)} \right. 
\nonumber \\[4pt]
&&\left. \times \left( \frac {A r_{lr}^{(0)} s^{(0)}_{,r} \delta r_{lr} + A r_{lr}^{(0)} \delta s}{1 - 2 A s^{(0)}} - \delta r_{lr} \right) \right],
\end{eqnarray}
with all the functions on the right hand side, together with their derivatives, evaluated at the commutative value $r = r_{lr}^{(0)}$.

%\bibliography{refart}

\end{document}